\documentclass[12pt]{revtex4}
\usepackage[cp1251]{inputenc}
\usepackage{amsmath}
\usepackage{amsbsy}

\begin{document}

\title{Pseudogap state of underdoped cuprate HTSC \\
as a display of the Jahn-Teller pseudoeffect}
\author{G.G. Sergeeva, A.A. Soroka}
\affiliation{National Science Center "Kharkov Institute of Physics
and Technology", Akademicheskaya st. 1, 61108 Kharkov, Ukraine}
\maketitle

Abstract:

For underdoped cuprate HTSC within the framework of the two-component model
of charge carries (holes and Zhang-Rice polarons) it is considered near
degenerate states of hole molecular orbitals of one-site cluster $%
Cu^{2+}O_{4}^{2-}$ and two-site cluster
$Cu^{2+}O_{4}^{2-}+Cu^{2+}O_{4}^{2-}$ in $CuO_{2}$ planes. It is
shown that the vibronic mixing of two near-degenerate electronic
states with near energies, separated by the energy barrier $\Delta
\varepsilon $, and mixed by Jahn-Teller $Q_{2}$ normal mode of
oxygen ions leads to the two-site Jahn-Teller polaron. The
supposition that the pseudogap state is a consequence of the
Jahn-Teller pseudoeffect and that the energy barrier $\Delta
\varepsilon $ is stipulated by magnetic interaction is  discussed.
It is shown that existence of two-site Jahn-Teller polarons is
consistent with a recent observation of broken time-reversal
symmetry in the pseudogap state.


\strut

The Jahn-Teller pseudoeffect at first was predicted by U. Opic and M.H.L.
Pryce in 1957~\cite{OpicPryce}. In \cite{OpicPryce} it was shown that a
degeneracy between different electronic states leads 
to the Jahn-Teller effect~\cite{JT} and so the vibronic mixing of
near-degenerate electronic states leads to the dynamical
Jahn-Teller (JT) effect. At a later date the consequences of this
vibronic mixing of near-degenerate electronic states were named as
the JT pseudoeffect~\cite {OpicPryce,Longuet-Higgins,Bersuker}.
Actually, let us consider two non-degenerate electronic states
with near energies $\varepsilon _{\alpha },\varepsilon _{i}$,
separated by the gap $\Delta \varepsilon =\varepsilon _{\alpha
}-\varepsilon _{i}<\varepsilon _{j},j=\alpha ,i$, and being mixed
by vibronic (i.e. combined electron-vibration) interaction. At
describing the ground state the degeneracy $\Delta \varepsilon
\rightarrow 0$ occurs only in the vicinity of Fermi surface, that
leads to vibronic instability and to double-valued chemical
potential function.

The first theoretical attempt to connect low-temperature superconductivity
with the JT pseudoeffect was made in 1961 by R.K. Nesbet~\cite{Nesbet} who
calculated corrections to the Born-Oppenheimer approximation for
near-degenerate electronic states in the presence of strong enough
electron-phonon interaction. Obtained in \cite{Nesbet} electron
configurations differ qualitatively from Fermi-distribution of the normal
metal, and in the chemical potential function there are relative maximum
inside the Fermi surface and relative minimum just outside.

This work was published at the time of a triumph of the BCS theory and
didn't attract marked interest. It should be noted the work~\cite{Englman}
containing critical remarks to~\cite{Nesbet} that cannot be accepted.
Firstly, in~\cite{Englman} the electron and nuclear Hamiltonians are treated
independently, whereas a considering of nuclear motion in systems with
strong electron-phonon interaction, having low-lying electronic states
(active electrons) close to degeneracy, results in a transformation of the
two independent Hamiltonians of nuclei and active electrons to the combined
electron-nuclear Hamiltonian. That just as in the JT effect leads to a
splitting of energy levels of electrons on near degenerate orbitals, i.e. to
the dynamical JT effect. Secondly, in~\cite{Englman} the conclusion that an
interaction between the phonon and electron subsystems brings to a
considerable phonon contribution to the energy correction was obtained
within the Frohlich theory~\cite{Frholich}. But it is known that within the
framework of the Frohlich theory it is impossible to describe the phonon
spectrum correctly~\cite{Brovman-Kagan,Geilikman}.

Though at present, 40 years later, we may say that in
\cite{Nesbet} it was predicted the pseudogap state (a dynamical
analog of charge ordering) which is observed in doped
antiferromagnets (AFs) whose Fermi surface consists of electron
and hole regions almost coinciding under the translation over the
wave vector of an antiferromagnetic cell (dispersion laws with
nesting). Pseudogap (PG) state in doped antiferromagnets is a
state with the two types of excitations: the holes are light
carriers and heavy carriers are Zhang-Rice (ZR)
polarons~\cite{Zhang-Rice,Muller,Cooper_Uchida,Mihailovich}. For
cuprate HTSC the PG state precedes a transition to the
superconducting state: with decreasing temperature at $T_{c}<T\leq
T^{*}$ a maximum and a minimum are observed in the density of
states near Fermi level at $\,\varepsilon \sim \varepsilon _{F}$,
which are characteristic for the JT
pseudoeffect~\cite{OpicPryce,Nesbet}.

In the present paper for underdoped cuprate HTSC within the framework of the
two-component model of charge carries~\cite{Muller} (which are holes and ZR
polarons) it is considered near degenerate states of hole molecular orbitals
of one-site and two-site clusters in $CuO_{2}$ planes. The first is an
occupied by a hole molecular orbital $\phi _{i}$ of ZR polaron which is a
hole localized on a square $Cu^{2+}O_{4}^{2-}$ with JT distortions by $Q_{2}$
normal mode of oxygen ions. Another molecular orbital is an unoccupied hole
molecular orbital $\phi _{\alpha }$ around the two-site cluster $%
Cu^{2+}O_{4}^{2-}+Cu^{2+}O_{4}^{2-}$ with common oxygen ion and
with antiferromagnetic spin ordered JT $Cu^{2+}$
ions~\cite{Sergeeva-Vakula}. JT distortions that lead to two-site
clusters come to an agreement with evidences of dynamical lattice
distortions on time scales $10^{-13}$ -- $10^{-15}$
s.~\cite{Kochelaev,Mook-Dogan,McQeeney}. Taking into account that
the vibronic mixing of these near-degenerate electronic states
leads to the dynamical JT effect in this paper the model of
pseudogap state as a consequence of the Jahn-Teller pseudoeffect
is discussed.

In the absence of the Jahn-Teller pseudoeffect (i.e. in the absence of near
degenerate electronic levels) the Born-Oppenheimer approximation is valid
for eigen-functions of the electron-nuclear Hamiltonian:
\begin{equation}
\Psi =\psi _{n}(q,Q)\,\varphi (n,Q),  \label{1}
\end{equation}
where $\psi _{n}(q,Q)$ and $\varphi (n,Q)$ are electronic and nuclear wave
functions. In (\ref{1}) $\psi _{n}(q,Q)$ is a solution of the Schredinger
equation for electrons at nuclei of the site $Cu^{3+}O_{4}^{2-}$ being in a
fixed position, and $\varphi (n,Q)$ is a solution of the Schredinger
equation for nuclei with electron energy $E_{n}(Q)$.

For JT $Cu^{2+}$ ions near-degenerate electronic states $\psi
_{n}(q,Q)$ depend on $Q_{2}$ JT normal mode, which leads to JT
distortion of squares $Cu^{2+}O_{4}^{2-}.$ Let us consider sixteen
one-site clusters from which there are nine $Cu^{2+}O_{4}^{2-}$
clusters (see Fig. 1). The occupied by a hole molecular orbital
$\phi _{i}$ is localized on one-site cluster $m_{ij}$ and forms
the ZR polaron (here the label $i$ is a number of horizontal
$Cu-O$ row, the label $j$ is a number of vertical $Cu-O$ row).
Taking into account the JT distortions for twelve $O^{2-}$ ions of
four nearest one-site clusters $m_{i+1,j}$, $m_{i-1,j},$
$m_{i,j+1}$, $m_{i,j-1}$ we can suppose that a hole can be excited
from the one-site molecular orbital $\phi _{i}$ of
$Cu^{2+}O_{4}^{2-}$ cluster with energy $\varepsilon _{i}$ to the
unoccupied two-site molecular orbital $\phi _{\alpha }$ of
$Cu^{2+}O_{4}^{2-} $+$Cu^{2+}O_{4}^{2-}$ cluster with energy
$\varepsilon _{\alpha }$ and with antiferromagnetic spin ordering
JT $Cu^{2+}$ ions. The hole excitation can be described as a
scattering of a charge carrier with absorption of the phonon
$\varepsilon _{\alpha }-\varepsilon _{i}=\hbar \omega _{k}$, where
$\omega _{k}$ is the frequency of $Q_{2}$ normal mode. At that
rate the electronic wave function $\psi _{n}(q,Q)$ is equal to
\begin{equation}
\begin{array}{cc}
\psi _{n}(q,Q)=\Phi _{0}+\sum_{Q_{k},i,\alpha }\Phi _{i}^{\alpha
}(is_{\alpha i}Q_{k}) &  \\
s_{\alpha i}=i\,(F_{i}^{\alpha })_{k}\left/ (\varepsilon _{\alpha
}-\varepsilon _{i})\right. ;\quad (F_{i}^{\alpha })_{k}=\langle \alpha
\,|\,\partial V/\partial Q_{k}\,|\,i\,\rangle  &
\end{array}
,  \label{2}
\end{equation}
where $\Phi _{0}$ is the Hartree-Fock many-electron ground state wave
function, and $\Phi _{i}^{\alpha }$ is the Slater determinant obtained from $%
\Phi _{0}$ by replacing an occupied hole molecular orbital $\phi _{i}$ by an
unoccupied hole molecular orbital $\phi _{\alpha }$.

Let us consider that after transition the occupied two-site
molecular orbital $\phi _{\alpha }$ belongs to two-site cluster
$m_{ij}+m_{i,j+1}$ (Fig. 1). The excited state of a hole which
occupies the two-site molecular orbital $\phi _{\alpha }$ can be
considered as a quasilocal state of a hole, i.e. the two-site
Jahn-Teller (TSJT) polaron~\cite{Sergeeva-Vakula}. TSJT polaron
moves with low damping in $CuO_{2}$ plane from two-site cluster
$m_{i,j}+m_{i,j+1}$ with two JT $Cu^{+2}$ ions to nearest two-site
clusters with antiferromagnetic spin ordered JT $Cu^{2+}$ ions. As
we see in Fig. 1,
any of unoccupied molecular orbitals of two-site clusters with two JT $%
Cu^{+2}$ ions can be occupied by a hole, for example $m_{i,j}+m_{i+1,j}$ , $%
m_{i,j+1}+m_{i+1,j+1}$, $m_{i,j}+m_{i,j-1}$, $m_{i,j}+m_{i-1,j}$ , $%
m_{i+1,j}+m_{i+1,j+1}$ . But the molecular orbital of two-site cluster $%
m_{i+1,j}+m_{i+1,j-1}$ cannot be occupied by a hole because it has only one
JT $Cu^{2+}$ ion. It should be noted that TSJT polaron cannot move along the
diagonal direction of squares formed by four $Cu^{2+}$ ions because there is
no oxygen ion on the diagonal between two $Cu^{2+}$ ions. For example, TSJT
from cluster $m_{ij}+m_{i,j+1}$ cannot be transferred on two-site cluster $%
m_{i,j+1}+m_{i+1,j}$ (Fig. 1).

In this model it is easy to illustrate the formation of the three spin
polaron~\cite{Kochelaev,Bianconi}. For $Cu^{3+}$ ion (site $m_{i-1,j+1})$,
which is surrounded by four $Cu^{2+}$ ions, positions of all nearest oxygen
ions are distorted by JT normal modes. It was shown by E.L. Nagaev~\cite
{Nagaev} that in layered AF charged impurity (here it is $Cu^{3+}$ ion)
leads to an almost spherical ferromagnetic cluster with the localized state
of a hole. So, in site $m_{i-1,j+1}\,$ the transition $Cu^{3+}\rightarrow
Cu_{\downarrow }^{2+}+h_{\uparrow }^{+}$ (with zero total spin) generates
the three spin polaron on two-site cluster with the localized state of a
hole (spin up) and parallel spins (down) of ions $Cu^{2+}$, and with three
dimensional JT distortions by $Q_{4}$ and $Q_{5}$ normal modes of seven
in-plane and four apex oxygen ions. This three spin polaron can be localized
by one of three two-site clusters around site $m_{i-1,j+1}$, for example by
cluster $m_{i-1,j+1}+m_{i-2,j+1}.$ The chains of three spin polarons form in
$CuO_{2}$ planes narrow stripes with distorted low temperature
tetragonal-like lattice~\cite{Bianconi}. But the transition $%
Cu^{3+}\rightarrow Cu_{\uparrow }^{2+}+h_{\downarrow }^{+}$ can generate the
TSJT polaron with the quasilocal state of a hole (spin down) and conserves
antiferromagnetic order in two-site cluster $m_{i-1,j+1}+m_{i-2,j+1}$.

For the ground state such near degeneracy only occurs at Fermi energy. In~%
\cite{Nesbet} it was obtained a relation for the transition matrix element
between an occupied orbital $\phi _{i}$ and an unoccupied orbital $\phi
_{\alpha }$ accompanied with absorption of the phonon with energy $%
\varepsilon _{\alpha }-\varepsilon _{i}=\hbar \omega _{k}$. For processes of
phonon absorption which occur between non-degenerate, but close to
degeneracy Born-Oppenheimer states the transition probability is proportion
to the quantity
\begin{equation}
\lim_{\delta _{i}\rightarrow 0}\frac{\delta _{i}}{(\varepsilon _{\alpha
}-\varepsilon _{i})^{2}+\delta _{i}^{2}}=\pi \,\delta (\varepsilon _{\alpha
}-\varepsilon _{i}),  \label{3}
\end{equation}
that brings to modification of stationary states on the different sides of
Fermi surface. Here $\delta _{i}=(s/v)\varepsilon _{i}$, where $s$ is the
velocity of longitudinal acoustic waves, $v$ is the velocity of electrons on
the Fermi surface. In consideration of (\ref{3}) and finiteness of the
quantity $\delta _{i}$ the correction $\varepsilon ^{\prime }(\sigma
_{i})=\mu (\sigma _{i})-\varepsilon (\sigma _{i})$ to the Hartree-Fock
energy of a single-electron state $\sigma _{i}$ has opposite signs on the
different sides of Fermi surface and is of the order~\cite{Nesbet}
\begin{equation}
\varepsilon ^{\prime }(\sigma _{F}^{\pm })\approx \pm (s/v)\varepsilon
_{F},\quad (s/v)\cong (Zm/3M)^{1/2},  \label{4}
\end{equation}
where $m$ is the electron mass, $M$ is the ion mass, $Z$ is the
ion charge, $\mu (\sigma _{i})$ is the electron chemical potential
function. As seen from (\ref{4}), there always exists a state
whose energy doesn't shift, and if value of $\varepsilon ^{\prime
}(\sigma _{F})$ is large enough, $\mu (\sigma _{i})$ may have
maximum on the inner side of Fermi surface and minimum on the
outside. Existence of such succession of maximum and minimum for
the electron chemical potential, as Frohlich pointed
out~\cite{Frholich}, implies presence of qualitative change in the
ground state of the electron system and can be a sign of the
superconducting state.

The series of optical
measurements~\cite{Cooper_Uchida,Mihailovich} have great
importance for studying the excitation of polaron into delocalized
states. At introducing doping into $CuO_{2}$ plane the optical
conductivity $\sigma (\omega )$ shows two peaks at 0.07 -- 0.1 eV
and at 0.8 -- 1 eV~\cite {Cooper_Uchida}, and
in~\cite{Mihailovich} was proposed that this structure of $\sigma
(\omega )$ corresponds to the transition of a polaron into
delocalized states with energy barrier of the order of the
Frank-Condon shift. Some of the first results which give evidence
for existence of the energy barrier for the polaron transport in
2D case were received in~\cite {Kabanov-Mashtakov}, where it was
shown that this barrier is attributed to the finite charge
carriers bandwidth.

The first conclusive proof of existence of TSJT polarons in
pseudogap and superconducting states was the observation at
$T<T^{*}$ in YBCO films of the double bimagnon-assisted absorption
band with maxima at $E_{1}=2.15$ eV and $E_{2}=2.28$ eV in
metallic films~\cite{Eremenko}. The first component peaked at
$E_{1}$ arises from the interband transition of ZR polaron at the
absorption of magnons for AF dielectric films. The observation at
$T<T^{*}$ of two components for metallic films evidences about
two-site nature of heavy charge carriers with the
antiferromagnetic core and the exchange energy
$E_{2}-E_{1}=J\approx $ 0.13 eV. This allows us to suppose that
the energy barrier
$\Delta\varepsilon=\varepsilon_{\alpha}-\varepsilon_{i}\simeq J$
for the hole excitation from the orbital $\phi_{i}$ of ZR polaron
to the unoccupied hole molecular orbital $\phi_{\alpha }$ around a
two-site cluster is stipulated by magnetic interaction between the
hole spin and the $Cu^{2+}$ spin: this magnetic interaction tries
to conserve antiparallel direction of these spins in each site of
TSJT polaron.

Recently T.D. Stanesku and P. Phillips~\cite{Stanesku-Phillips} showed that
origin of the pseudogap in all 2D doped Mott insulators is caused by local
correlations, and the energy barrier for the hole transport is $\sim t^{2}/U$%
, where $t$ is the nearest-neighbour hopping and $U$ is on-site
Coulomb repulsion. In the dynamic 2D Hubbard model they computed
the density of states and found that near Fermi level at
$\varepsilon <\varepsilon _{F}$ maximum is observed and minimum is
observed at $\varepsilon >\varepsilon _{F} $. This means that the
pseudogap reflects the restricted phase space where strongly
correlated excitations on neighbouring sites in are taken into
account. It should be noted that received in
\cite{Stanesku-Phillips} estimation coincides with measurement
\cite{Eremenko} of the energy barrier $\Delta \varepsilon \simeq
J\sim t^{2}/U$.

Thus, in $CuO_{2}$ plane two near-degenerate electronic states with near
energies $\varepsilon _{\alpha },\varepsilon _{i}$, separated by the gap $%
\Delta \varepsilon $ and mixed by JT $Q_{2}$ normal mode of oxygen
ions, lead to transition of the system to the excited state with
new quasi-particles --- ''heavy holes'' with two-site
antiferromagnetic core which may be considered as two-site
Jahn-Teller polarons~\cite {Sergeeva-Vakula,Mook-Dogan,McQeeney}.
It is discussed the supposition that the pseudogap state is a
consequence of the Jahn-Teller pseudoeffect, and the energy
barrier $\Delta \varepsilon $ is stipulated by magnetic
interaction. Two-site JT polarons move in $CuO_{2}$ plane along
the $(\pi ,0) $ and $(0,\pi )$ directions and not along $(\pi ,\pi
)$. Recent circular dichroism experiments~\cite{Kaminski} point to
time-reversal as the relevant symmetry that is broken in PG state
along the $(\pi ,0)$ and $(0,\pi )$ directions and not along $(\pi
,\pi )$. In the context of this paper we can consider the movement
of two site JT polarons (which cannot move along $(\pi ,\pi )$
direction) as local currents that lead to breaking of
time-reversal symmetry in PG state. This symmetry must be broken
as well in the superconducting state because there are
observations of pseudogap at $T<T_{c}.$

\pagebreak

FIGURE CAPTIONS
\begin{figure}[h!]
\caption{Transition from the one-site Zhang-Rice polaron to the
two-site Jahn-Teller polaron in $CuO_{2}$ plane. The label $i$ is
a number of horizontal $Cu-O$ row, $j$ is a number of vertical
$Cu-O$ row. Black circles denote $Cu^{3+}$ ions, black circles
with arrows denote $Cu^{2+}$ ions (where arrows specify the
direction of spin of these ions), open circles denote $O^{2-}$
ions, the small open circle with an arrow denotes a hole $h^+ $.
Small shading inside one-site cluster $m_{ij}$ denotes the
occupied one-site hole molecular orbital $\phi_i$ of the
Zhang-Rice polaron. Big shading outside two-site cluster
$m_{ij}+m_{i,j+1}$ denotes the occupied two-site hole molecular
orbital $\phi_{\alpha}$ of the Jahn-Teller polaron.} \label{fig1}
\end{figure}

\end{document}